\documentclass[12pt]{iopart}

\usepackage{graphicx}
\usepackage{subfigure}
\usepackage{color}

\newcommand{\bra}[1]{\langle #1 |}
\newcommand{\ket}[1]{| #1 \rangle}

\newcommand{\const}[0]{ {\rm const}}

\newcommand{\eqref}[1]{(\ref{#1})}

\usepackage{graphicx}
\usepackage{subfigure}
\usepackage{color}

\begin{document}

\title[  ]{Spectral gap and the exponential localization in general one-particle systems}

\author{Tomotaka Kuwahara}

\address{Department of Physics, The University of Tokyo, Komaba, Meguro, Tokyo 153-8505}
\ead{tomotaka@iis.u-tokyo.ac.jp}

\begin{abstract}
We investigate the relationship between the spectral gap $\delta E_0$ and the localization length $\xi$ in general one-particle systems.
A relationship for many-body systems between the spectral gap and the exponential clustering has been derived from the Lieb-Robinson bound, which reduces to the inequality $\xi \le \const. \times \delta E_0^{-1} $ for one-particle systems. This inequality, however, turned out not to be optimal qualitatively.
As a refined upper bound, we here prove the inequality $\xi \le \const. \times \delta E_0^{-1/2}$  in general one-particle systems.
Our proof is not based on the Lieb-Robinson bound, but on our complementary inequality related to the uncertainty principle [T. Kuwahara, J. Phys. A: Math. Theor. {\bf46} (2013)].
We give a specific form of the upper bound and test its tightness in the tight-binding Hamiltonian with a diagonal impurity, where the localization length behaves as $\xi \sim \delta E_0^{-1/2}$.
We ensure that our upper bound is quantitatively tight in the case of nearest-neighbor hopping.

\end{abstract}

\maketitle

\section{Introduction}

In quantum many-body systems, the spectral gap plays important roles in determining the fundamental properties of the ground state; we here define the spectral gap, denoted by $\delta E_0$, as the energy difference between the ground state and the first excited state.
In analyzing the ground state's properties in terms of the spectral gap, the Lieb-Robinson~\cite{LR_bound} bound has been one of the strongest tools; it characterizes the causality in non-relativistic systems~\cite{Nachtergaele3,Bravyi2,Osborne,Hamma,Nachtergaele4,Schwarz,Cheneau}. 
In other words, it bounds the velocity of the information transfer.
There are many results which come from the Lieb-Robinson bound~\cite{Hastings,Nachtergaele2,Hastings4,Hastings5,Eisert,Gottesman,Hastings6}. 
One of the most important results is a proof of the following folk theorem~\cite{Hastings,Hastings2,Hastings3,Nachtergaele,Koma}: the non-vanishing spectral gap implies the exponential decay of the spatial correlations. 
Such a theorem on the exponential clustering gives an inequality for the correlation length $\xi_{\rm cor}$ in the form $\xi_{\rm cor}\le \const. \times \delta E_0^{-1}$.
This inequality is tight up to a constant coefficient; if we consider quantum critical systems, this inequality indicates $z\ge 1$, where $z$ is the dynamical critical exponent. 
We indeed know that in many systems the dynamical critical exponent~$z$ is equal to unity~\cite{Vojtan,BookQP}.

The above inequality is also applied to one-particle systems, which are the simplest instances of quantum many-body systems.
For one-particle systems, the inequality of the exponential clustering gives the upper bound of the localization length~$\xi$ instead of the correlation length~$\xi_{\rm cor}$, namely $\xi \le \const. \times \delta E_0^{-1}$.
This inequality, however, does not reflect the empirical property of the ground state in one-particle systems.
We expect that the localization length in the ground state behaves as $\xi\sim \delta E_0^{-1/2}$, as in the tight-binding Hamiltonian with a diagonal defect~\cite{Apollaro} for example.
We thereby expect that the inequality of the exponential clustering should have the possibility to be refined for one-particle systems.

In the present paper, we make this empirical expectation rigorous; we prove $\xi \le \const. \times \delta E_0^{-1/2}$ mathematically for general one-particle systems. 
The proof of this inequality is not based on the Lieb-Robinson bound, but on our inequality derived in Ref.~\cite{Kuwahara}.
The inequality gives the complementary relationship~$\delta E_0 \cdot (\Delta x)^2 \le \const.$, where $\Delta x$ is the fluctuation of the position with respect to the ground state.
This is qualitatively explained by the uncertainty principle as follows.
Let us consider a one-dimensional square-well model.  In this model, we expect the spectral gap $\delta E_0$ as $\delta E_0\sim (\Delta p)^2$, and we infer $ \Delta x \sim \const. \times (\delta E_0)^{-1/2}$ from the uncertainty relationship  $\Delta x \cdot  \Delta p \sim \hbar$; such an inference turns out to be rigorous as was proved in Ref.~\cite{Kuwahara}.
This relationship, however, only refers to the variance of the position and is insufficient to determine that the localization length also satisfies the inequality~$\xi \le \const. \times \delta E_0^{-1/2}$.
We therefore refine the original proof and derive the upper bound of the localization length.

We here show the outline of the present paper.
The main result is a derivation of the refined inequality between the spectral gap~$\delta E_0$ and the localization length~$\xi$ in general one-particle systems. 
In Section~2, we first show the fundamental setup of the system, definition of the terms, and the original complementary inequality in Ref.~\cite{Kuwahara}.
In Section~3, we review the relationship between the spectral gap and the localization.
We refer to the Chebyshev inequality in Section~3.1 and the Lieb-Robinson bound in Section~3.2, respectively.
We show our main inequality in Section~4.1 and discuss the tightness of the inequalities in Section~4.2.
In Section~5, discussion concludes the paper.


\section{Statement of the problem}
We first define the general one-particle Hamiltonian: 
\begin{eqnarray}
 H=\sum_{x, x'=1}^{L}\sum_{i,j=1}^{N_0} (  h_{(x,i),(x',j)} a_{(x,i)}^{\dagger} a_{(x',j)} + {\rm h.c.}    ) ,\label{Hamiltonian_tight_binding}
\end{eqnarray}
where $a_{(x,i)}^{\dagger}$ and $a_{(x,i)}$ are the creation and annihilation operators on the site $(x,i)$, respectively, and $\{i\}_{i=1}^{N_0}$ are internal parameters which is assigned to each site $x$.
Note that this includes the Hamiltonian in high-dimensional systems.
We assume an exponential decay of the hopping rate from the site~$x$ to $x'$ $(x\neq x')$, namely,
\begin{eqnarray}
\biggl \|\sum_{i,j=1}^{N_0}  (  h_{(x,i),(x',j)} a_{(x,i)}^{\dagger} a_{(x',j)} + {\rm h.c.}    )   \biggr \|_2 \le \mathcal{V}(x-x') \equiv  C_v e^{-\mu |x-x'|}, \label{Condition_for_interaction}
\end{eqnarray}
where $\|\cdots\|_2$ denotes the spectral norm.
Note that we do not have \textit{any} restrictions to the parameters $\{h_{(x,i),(x,j)}\}$.

Second, we denote the ground state $\ket{\psi_0}$ of the Hamiltonian as 
\begin{eqnarray}
\ket{\psi_0}\equiv\sum_{x=1}^{L}\sum_{i=1}^{N_0} \alpha_{(x,i)} \ket{x,i}     ,   \label{Psi_Ground_state} 
\end{eqnarray}
where $\{\alpha_{(x,i)}\}_{(x,i)}$ are complex coefficients and the bases $\{\ket{x,i}\}_{(x,i)}$ are defined as $\ket{x,i}\equiv a_{(x,i)}^\dagger \ket{{\rm vac}}$
with $\ket{{\rm vac}}$ denoting the vacuum state.

Third, we denote the eigenenergies of the Hamiltonian $H$ as $\{E_n\}$ in non-descending order $(E_0\le E_1\le E_2\cdots)$.
We here do not consider the case where the ground state is degenerate and define the spectral gap $\delta E_0$ as 
\begin{eqnarray}
\delta E_0 \equiv  E_1-E_0 > 0. \label{Definition_of_energy_gap}
\end{eqnarray}
Using the result in Ref.~\cite{Kuwahara}, we can obtain the following inequality complementary between the spectral gap and the fluctuation:
\begin{eqnarray}
\delta E_0\cdot (\Delta G)^2 \le  \frac{ | \bra{\psi_0} H_{{\rm OD}}   \ket{\psi_0}|}{2}  \label{New_Final_form_of_the_inequality} 
\end{eqnarray}
with 
\begin{eqnarray}
&G\equiv \sum_{x=1}^L g(x) \sum_{i=1}^{N_0} \ket{x,i} \bra{x,i} ,\label{G_tight_binding} \\
&H_{{\rm OD}} \equiv \sum_{x,x'=1}^{L}\bigl[ g(x)-g(x') \bigr]^2  \sum_{i,j}^{N_0} (  h_{(x,i),(x',j)} a_{(x,i)}^{\dagger} a_{(x',j)} + {\rm h.c.}    ) ,\label{definition_of_HOD}
\end{eqnarray}
where $g(x)$ is an arbitrary real function and $(\Delta G)^2$ is the variance of the operator $G$ with respect to the ground state: 
\begin{eqnarray}
(\Delta G)^2 &\equiv \bra{\psi_0} G^2 \ket{\psi_0}-\bigl(\bra{\psi_0} G  \ket{\psi_0}\bigr)^2 , \quad \Delta G\ge 0.
\end{eqnarray}
Note that the terms in $H_{{\rm OD}}$ which contain $\{h_{(x,i),(x,j)}\}$ always vanish because of the term $\bigl[ g(x)-g(x') \bigr]^2$; this is the reason why we do not need to have any restrictions to the parameters $\{h_{(x,i),(x,j)}\}$.
Since the expression of $H_{\rm OD}$ had not been explicitly given in Ref.~\cite{Kuwahara} for the general one-particle Hamiltonian~\eqref{Hamiltonian_tight_binding}; we show the proof in Appendix~A. 
The inequality~\eqref{New_Final_form_of_the_inequality} indicates that the spectral gap is bounded from above by the fluctuation of $G$ and vice versa.

In the following, we define the distribution of the particle density as 
\begin{eqnarray}
p_x \equiv \sum_{i=1}^{N_0} |\alpha_{(x,i)}|^2 \quad {\rm and} \quad 
P(|x-\langle x\rangle| \ge  R) \equiv \sum_{|x-\langle x\rangle| \ge R}  p_x , \label{definition_of_the_density}
\end{eqnarray}
where $\langle x \rangle$ is the average position of the particle, namely $\langle x \rangle \equiv \bra{\psi_0} X  \ket{\psi_0}$ with
\begin{eqnarray}
&X \equiv \sum_{x=1}^L x\sum_{i=1}^{N_0} \ket{x,i} \bra{x,i}. \label{X_tight_binding} 
\end{eqnarray}
We investigate the relationship between the decay law of $P(|x-\langle x\rangle| \ge  R)$ and the spectral gap~$\delta E_0$.


\section{Relationship between the Decay law and the spectral gap}

\subsection{Decay law by the Chebyshev inequality}
We first discuss the decay law using the Chebyshev inequality, which gives a power law decay with respect to the spectral gap.
This decay law utilizes only the upper bound of the variance $(\Delta X)^2$.

First, the Chebyshev inequality is given by
\begin{eqnarray}
P(|x-\langle x\rangle| \ge  k \Delta X ) \le \frac{1}{k^2},
\end{eqnarray}
where $\Delta X$ is the standard variation of the position operator $X$. 
By letting $k=R/\Delta X$, we have 
\begin{eqnarray}
P(|x-\langle x \rangle| \ge  R ) \le \frac{(\Delta X)^2}{R^2}.   \label{Trivial_bound2}
\end{eqnarray}
We here relate $\Delta X$ to the spectral gap $\delta E_0$ using of the inequality~\eqref{New_Final_form_of_the_inequality}. 

In order to derive the relation, we first let $g(x)=x$ in Eq.~\eqref{G_tight_binding} and obtain
\begin{eqnarray}
\fl | \bra{\psi_0} H_{{\rm OD}}   \ket{\psi_0}| &= \sum_{x, x'=1}^L(x-x' )^2 \sum_{i,j=1}^{N_0} (  h_{(x,i),(x',j)} \alpha_{(x,i)}^\ast \alpha_{x',j} +{\rm c.c.} )  \nonumber \\
\fl &\le \sum_{x, x'=1}^L(x-x' )^2  \biggl \|\sum_{i,j=1}^{N_0}  (  h_{(x,i),(x',j)} a_{(x,i)}^{\dagger} a_{(x',j)} + {\rm h.c.}    )   \biggr \|_2  (p_{x} + p_{x'} )   \nonumber \\ 
\fl &\le \sum_{x, x'=1}^L(x-x' )^2 \mathcal{V}(x-x')   (p_{x} + p_{x'} )  \nonumber \\
\fl &=2\sum_{x, x'=1}^L (x-x' )^2 \mathcal{V}(x-x')   p_{x}  \equiv \sum_{x=1}^L \mathcal{V}_x p_{x}  ,  \label{Inequality_chebyshef}
\end{eqnarray}
where we define $\mathcal{V}_x \equiv 2 \sum_{x'=1}^{L}\mathcal{V}(x-x') (x-x' )^2$.
From the first line to the second line, we utilize the following inequality:
\begin{eqnarray}
&\sum_{i,j=1}^{N_0} (  h_{(x,i),(x',j)} \alpha_{(x,i)}^\ast \alpha_{x',j} +{\rm c.c.} )  \nonumber \\
 =&(p_{x} + p_{x'} ) \biggl \langle \psi_{0}^{x,x'} \biggl| \sum_{i,j=1}^{N_0}  (  h_{(x,i),(x',j)} a_{(x,i)}^{\dagger} a_{(x',j)} + {\rm h.c.}    ) \biggr | \psi_0^{x,x'}\biggr \rangle \nonumber \\
 \le &(p_{x} + p_{x'} )  \biggl \|\sum_{i,j=1}^{N_0}  (  h_{(x,i),(x',j)} a_{(x,i)}^{\dagger} a_{(x',j)} + {\rm h.c.}    )   \biggr \|_2
\end{eqnarray}
with 
\begin{eqnarray}
\ket{\psi_{0}^{x,x'}}\equiv  \frac{1}{\sqrt{p_{x} + p_{x'}}} \sum_{i=1}^{N_0} ( \alpha_{(x,i)} \ket{x,i}+  \alpha_{(x',i)} \ket{x',i})  ,
\end{eqnarray}
where $1/\sqrt{p_{x} + p_{x'}}$ is the normalization factor.


From the inequalities~\eqref{New_Final_form_of_the_inequality} and \eqref{Inequality_chebyshef}, we have 
\begin{eqnarray}
(\Delta X)^2 \le  \frac{\max_{x} (\mathcal{V}_x) }{2\delta E_0}  , \label{Upper_bound_of_the_variance_of_PositionX}
\end{eqnarray}
where $\max_{x} (\mathcal{V}_x)$ is the maximum value of $\mathcal{V}_x$ among $x=1,2,\ldots,L$.
By applying the above inequality to \eqref{Trivial_bound2}, we obtain
\begin{eqnarray}
P(|x-\langle x \rangle| \ge  R ) \le  \frac{\max_{x} (\mathcal{V}_x)}{2R^2\delta E_0 } ,   \label{Trivial_bound}
\end{eqnarray}
which gives the power law decay of $R^{-2}$.
This upper bound is derived only from the information on the variance of the position operator.
In the derivation of the inequality~\eqref{Trivial_bound}, we used the position operator $X$ for the operator $G$ in \eqref{G_tight_binding} and utilized the inequality $(\Delta X)^2 \le \const. \times (\delta E_0)^{-1}$.
In fact, we can take the operator $G$ arbitrarily and derive a stronger upper bound by choosing the operator $G$ as in Section~\ref{sec_main_result}.

 
\subsection{Decay law by the Lieb-Robinson bound}
We here discuss the decay law in Refs.~\cite{Hastings,Hastings2,Hastings3,Nachtergaele,Koma}, which utilize the Lieb-Robinson bound. 
The Lieb-Robinson bound characterizes the velocity of the information transfer and is a strong tool to analyze the fundamental properties of the quantum many-body systems~\cite{Nachtergaele3,Bravyi2,Osborne,Hamma,Nachtergaele4,Schwarz,Cheneau}. 
In Refs.~\cite{Hastings,Hastings2,Hastings3,Nachtergaele,Koma}, they proved the exponential clustering for general quantum many body systems; these results can be also applied to the Hamiltonian~\eqref{Hamiltonian_tight_binding}.
We can obtain the upper bound 
\begin{eqnarray}
P(|x-\langle x \rangle| \ge  R ) \le  c_1 e^{-R/\xi_0},   \label{LR_derive_bound}
\end{eqnarray}
where
\begin{eqnarray}
 \xi_0 = c_2 +\frac{c_3}{\delta E_0}   \label{LR_derive_bound}
\end{eqnarray}
with $c_1$, $c_2$ and $c_3$ being constants which only depend on details of the Hamiltonian. 
This upper bound is qualitatively stronger than the one in the inequality \eqref{Trivial_bound}.
The inequality~\eqref{LR_derive_bound}, however, is not optimal as will be mentioned later; indeed, we empirically know that the localization length in the ground state is roughly proportional to $\delta E_0^{-1/2}$, not $\delta E_0^{-1}$.



\section{Main result} \label{sec_main_result}

\subsection{Main theorems}
We here prove that the localization length is lower than $O(\delta E_0^{-1/2})$ based on the inequality~\eqref{New_Final_form_of_the_inequality}. 
We now prove the following Theorem~1.

\textit{Theorem~1}.
When we assume 
\begin{eqnarray}
R \ge r_1\equiv \sqrt{\frac{2e+1}{1-s}} \Delta X   \quad {\rm for} \quad 0<s<1 ,\label{lower_bound_R}
\end{eqnarray}
the distribution $P(|x-\langle x \rangle| \ge  R )$ satisfies the following inequality:
\begin{eqnarray}
P(|x-\langle x \rangle| \ge  R ) \le \frac{2e(2-s)+1}{4(2e+1)}  \exp \Biggl(-\frac{R-r_1}{\xi_1} \Biggr) , \label{Final_form_of_the_inequality_decay}
\end{eqnarray}
where the localization length $\xi_1$ is given by
\begin{eqnarray}
\xi_1\equiv \max \Biggl(\frac{3}{2}\sqrt{\frac{ (4e^2+1) C_1}{es \delta E_0}}, \frac{3\ln (2e)}{\mu}\Biggr) .\label{the_way_of_choice_of_xi}
\end{eqnarray}
The parameter $C_1$ is a constant defined in Eq.~\eqref{the_definition_of_C1} below, which depends only on the values $C_v$ and $\mu$ in Eq.~\eqref{Condition_for_interaction}.


\textit{Comment}.
This inequality shows that the distribution $P(|x-\langle x \rangle| \ge  R )$ decays exponentially for $R\gg \Delta X$ and the localization length~$\xi_1$ is bounded from above by $O(\delta E_0^{-1/2})$.
Such an upper bound for the localization length is tight up to a constant coefficient.
Note that the localization length $\xi_1$ is characterized only by the hopping parameters  in Eq.~\eqref{Condition_for_interaction} and the spectral gap $\delta E_0$. 
As has been mentioned in Ref.~\cite{Kuwahara}, the information of the potential terms $\{h_{(x,i),(x,j)}\}$ in the Hamiltonian is contained in the spectral gap $\delta E_0$.
As for the coefficient of $\delta E_0^{-1/2}$, however, the upper bound by Eq.~\eqref{the_way_of_choice_of_xi} does not seem to be so tight as will be mentioned later.
If we restrict the Hamiltonian to the nearest-neighbor hopping, we can obtain a tighter upper bound as in Theorem~2 below.

 \begin{figure}
\centering
\includegraphics[clip, scale=0.8]{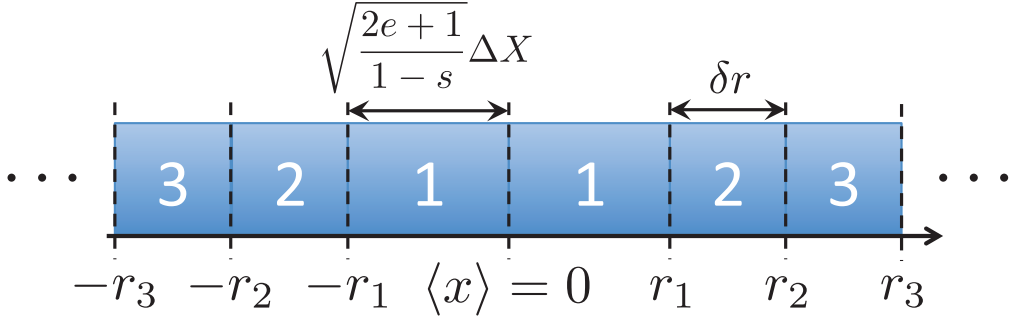}
\caption{We define the point where $x=\langle x \rangle$ as the origin $x=0$.
We then split the space to several regions; the first region has the width $r_1=\sqrt{\frac{2e+1}{1-s}} \Delta X$ and the other regions have the width $\delta r$. 
}
\label{fig:region_definition}
\end{figure}

\textit{Proof}.
We first take the origin as the point where 
\begin{eqnarray}
x=\langle x \rangle=0.
\end{eqnarray} 
Note that we consider a discrete space and each of the initial points may now be given by a non-integral number as
\begin{eqnarray}
-\langle x \rangle+1,\  -\langle x \rangle+2, \ \ldots, \ -\langle x \rangle+L, \label{non_integral_point_tight_binding}
\end{eqnarray} 
respectively.
We next split the coordinate into some regions with their width $\delta r$ (Figure.\ref{fig:region_definition}); we will properly choose the parameter $\delta r$ afterward.

The $n$th region is defined as
\begin{eqnarray}
 r_{n-1}\le  |x| < r_n, \label{each_region_definition}
\end{eqnarray}  
where we define $r_0=0$, $r_1=\sqrt{\frac{2e+1}{1-s}} \Delta X$ and 
\begin{eqnarray} 
r_n=r_{n-1} +\delta r  \label{each_region_definition_rn}
\end{eqnarray} 
for $n\ge 2$.
We define the region $N$ as the one where the point $x=R$ is included, which is given by
\begin{eqnarray}
N =\biggl \lceil \frac{R-r_1}{\delta r}+1  \biggr \rceil \label{The_Definition_of_Large_N}
\end{eqnarray} 
from the definition~\eqref{each_region_definition_rn}, where $\lceil \cdots \rceil$ denotes the ceiling function.
Note that the point $x=R$ satisfies 
\begin{eqnarray}
 r_{N-1} \le R <  r_N, \label{THe_point_of_R}
\end{eqnarray} 
and $N$ is larger than $2$ because $R \ge r_1$ from the assumption~\eqref{lower_bound_R}.

We define the cumulative probability distribution in the region $n$ as $P_n$:
\begin{eqnarray}
P_n = \sum_{ r_{n-1}\le  |x| < r_n} p_x , \label{THe_Definition_of_Large_P_n}
\end{eqnarray} 
where $p_x$ is defined in Eq.~\eqref{definition_of_the_density}.
We also define $P_{n\ge n_0}$ as
\begin{eqnarray}
P_{n\ge n_0} = \sum_{|x|\ge r_{n_0-1}} p_x , \label{THe_Definition_of_Large_P_n_ge_n_0}
\end{eqnarray} 
where $\sum_{x_1\le x \le x_2}$ denotes the discrete summation over the points \eqref{non_integral_point_tight_binding} in $x_1\le x \le x_2$.

As in the inequality~\eqref{THe_point_of_R} for $R$, the point $|x-\langle x \rangle|=  R$ is included in the $N$th region, and we have
\begin{eqnarray}
P(|x-\langle x \rangle| \ge  R ) \le P_{n\ge N}=\sum_{|x|\ge r_{N-1}} p_x . \label{P_x_R_and_P_n_N}
\end{eqnarray}
We then consider the upper bound of $P_{n\ge N} $ instead of $P(|x-\langle x \rangle| \ge  R )$.
In the following, we derive an upper bound of $P_{n\ge n_0}$ for arbitrary $n_0\ge 2$.

We here show the outline of the proof.
We first bound $P_{n\ge n_0+1}$ from above by the use of $P_2,P_3,\ldots,P_{n_0}$ as 
\begin{eqnarray}
P_{n\ge n_0+1} \le A (P_{n_0} + B P_{n_0-1}+ B^2 P_{n_0-2}+ \cdots B^{n_0-1} P_{1}  ), \label{General_inequality_P_n_01}
\end{eqnarray}
where the coefficients $A$ and $B$ can be determined by the spectral gap $\delta E_0$, the width of the region $\delta r$ and the hopping parameters in the inequality~\eqref{Condition_for_interaction}.
In order to obtain the specific forms of $A$ and $B$, which will be given in  Eq.~\eqref{choice_of_A_and_B} below, we utilize the inequality~\eqref{New_Final_form_of_the_inequality}; it is crucial to choose the operator $G$ properly.
We will finally prove that the inequality \eqref{General_inequality_P_n_01} reduces to the main inequality~\eqref{Final_form_of_the_inequality_decay} by choosing the parameter 
$\delta r$ equal to $\xi_1$ in Eq.~\eqref{the_way_of_choice_of_xi}.

 \begin{figure}
\centering
\includegraphics[clip, scale=0.6]{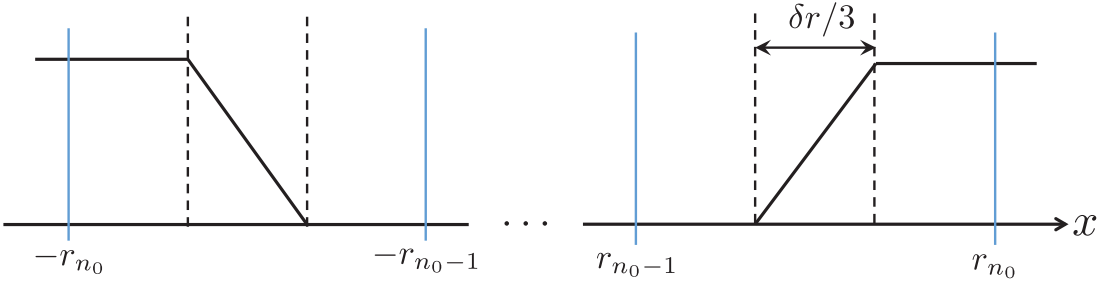}
\caption{The function $g(x)$ in operator $G$. From such a choice of $g(x)$, we can obtain the local information on the regions close to $n_0$, which is followed by the inequality~\eqref{General_inequality_P_n_01}. 
}
\label{fig:definition_of_the_function_gx}
\end{figure}

In order to derive the restrictions on $P_n$ and $P_{n\ge n_0}$, we choose the function $g(x)$ in Eq.~\eqref{definition_of_HOD} as follows (Figure.~\ref{fig:definition_of_the_function_gx}):
\begin{eqnarray}
\fl&g(x) = 
\cases{ 0\quad   {\rm for} \quad  |x|-r_{n_0-1}  \le  \delta r/3 ,    \\
|x| -(r_{n_0-1}+ \delta r/3 ) \quad  {\rm for} \quad  \delta r/3 \le |x|-r_{n_0-1}\le  2\delta r/3,  \\  
  \delta r/3 \quad   {\rm for} \quad  2\delta r/3 \le |x|-r_{n_0-1} .}   \label{choice_of_g_x_of_the_OpeRator_G}
\end{eqnarray}
Such a choice of $g(x)$ gives a qualitatively tight upper bound.
Using the above function $g(x)$, we calculate a lower bound of $\bra{\psi_0}G^2\ket{\psi_0}$ and upper bounds of $|\bra{\psi_0} H_{{\rm OD}} \ket{\psi_0}|$ and $\bra{\psi_0}G\ket{\psi_0}^2$, which are the elements in the inequality~\eqref{New_Final_form_of_the_inequality}.

We first calculate an upper bound of $|\bra{\psi_0} H_{{\rm OD}} \ket{\psi_0}|$, which is given by
\begin{eqnarray}
| \bra{\psi_0} H_{{\rm OD}}   \ket{\psi_0}| \le C_1P_{n_0}&+C_1\sum_{n\ge n_0+1} e^{-\mu ( (n-n_0) \delta r -2\delta r/3)} P_{n} \nonumber \\
&+C_1\sum_{n\le n_0-1} e^{-\mu ((n_0-n)\delta r -2\delta r/3)} P_{n},\label{last_form_of_the_inequality_lemma}
\end{eqnarray}
where $C_1$ is defined as
\begin{eqnarray}
C_1 \equiv 4C_v \sum_{x=0}^\infty (x+1)^2 e^{-\mu x} . \label{the_definition_of_C1}
\end{eqnarray}
We prove the inequality~\eqref{last_form_of_the_inequality_lemma} in Appendix~B.

Next, we calculate an upper bound of  $\bra{\psi_0}G\ket{\psi_0}^2$ and a lower bound of $\bra{\psi_0}G^2\ket{\psi_0}$, which are straightforwardly given by
\begin{eqnarray}
\bra{\psi_0}G\ket{\psi_0}^2 \le  (P_{n_0}+P_{n\ge n_0+1})^2 \biggl(\frac{\delta r}{3}\biggr)^2 \label{inequality_F_square_2}
\end{eqnarray} 
and 
\begin{eqnarray}
\bra{\psi_0}G^2\ket{\psi_0}\ge P_{n\ge n_0+1} \biggl(\frac{\delta r}{3}\biggr)^2,\label{inequality_F_square}
\end{eqnarray} 
respectively.

In the following, we consider the following two cases 1 and 2.\\
\textit{(case~1)}
\begin{eqnarray}
(1-s)\bra{\psi_0}G^2\ket{\psi_0}\le \bra{\psi_0}G\ket{\psi_0}^2 .  \label{G2_AG2_condition_case1}
\end{eqnarray}
\textit{(case~2)}
\begin{eqnarray}
(1-s)\bra{\psi_0}G^2\ket{\psi_0}> \bra{\psi_0}G\ket{\psi_0}^2  . \label{G2_AG2_condition_case2}
\end{eqnarray}

We first consider the case~1.
From the inequalities~\eqref{inequality_F_square_2} and \eqref{inequality_F_square}, we obtain
\begin{eqnarray}
\fl P_{n\ge n_0+1}\biggl(\frac{\delta r}{3}\biggr)^2\le \bra{\psi_0}G^2\ket{\psi_0}\le \frac{1}{1-s} \bra{\psi_0}G\ket{\psi_0}^2   \le \frac{1}{1-s}(P_{n_0}+P_{n\ge n_0+1})^2 \biggl(\frac{\delta r}{3}\biggr)^2,
\end{eqnarray}
which reduces to
\begin{eqnarray}
P_{n\ge n_0+1}\le  \frac{1}{1-s}(P_{n_0}+P_{n\ge n_0+1})^2.
\end{eqnarray}
We then utilize the Chebyshev inequality~\eqref{Trivial_bound2} as 
\begin{eqnarray}
P_{n_0}+P_{n\ge n_0+1}  \le  \frac{(\Delta X)^2}{r_1^2} =\frac{1-s}{2e+1}  \equiv (1-s) P_{0}, \label{1Definition_of_P_0}
\end{eqnarray}
because as in Figure~\ref{fig:region_definition} we have $|x| \ge r_1=\sqrt{\frac{2e+1}{1-s}}\Delta X$ in the $n_0$th region ($n_0\ge 2$).
Therefore, we have
\begin{eqnarray}
P_{n\ge n_0+1}   & \le P_0 (P_{n_0}+P_{n\ge n_0+1}) \nonumber \\
 P_{n\ge n_0+1} & \le \frac{P_0}{1-P_0}P_{n_0} = \frac{1}{2e} P_{n_0},   \label{Case_1_main_inequality34}
\end{eqnarray}
where $P_{0}=1/(2e+1)$ from the definition~\eqref{1Definition_of_P_0}.

We next consider the case~2.
From the inequalities~\eqref{inequality_F_square} and \eqref{G2_AG2_condition_case2}, we calculate the variance $(\Delta G)^2$ as
\begin{eqnarray}
(\Delta G)^2= \bra{\psi_0}G^2\ket{\psi_0} - \bra{\psi_0}G\ket{\psi_0}^2  > s \bra{\psi_0}G^2\ket{\psi_0} \ge s P_{n\ge n_0+1}  \biggl(\frac{\delta r}{3}\biggr)^2 .\label{last_form_of_the_inequality2_F}
\end{eqnarray}
From the inequalities \eqref{last_form_of_the_inequality_lemma} and \eqref{last_form_of_the_inequality2_F}, we calculate the fundamental inequality~\eqref{New_Final_form_of_the_inequality} as
\begin{eqnarray}
2s P_{n\ge n_0+1} \delta E_0 \biggl(\frac{\delta r}{3}\biggr)^2 \le  C_1P_{n_0}&+C_1\sum_{n\ge n_0+1} e^{-\mu ((n-n_0) \delta r -2\delta r/3)} P_{n} \nonumber \\
&+C_1\sum_{n\le n_0-1} e^{-\mu ((n_0-n)\delta r -2\delta r/3)} P_{n}.  \label{inequality_for_P_n_n_0_p1}
\end{eqnarray}
We then derive the upper bound of $P_{n\ge n_0+1}$ as in the inequality~\eqref{General_inequality_P_n_01} based on this inequality.

We first simplifies the right-hand side of the inequality~\eqref{inequality_for_P_n_n_0_p1} into
\begin{eqnarray}
C_1\sum_{n\ge n_0+1}e^{-\mu ((n-n_0) \delta r -2\delta r/3)}   P_{n}\le C_1e^{-\mu \delta r/3} P_{n\ge n_0+1}\label{1564Inequality_for_first00}
\end{eqnarray}
and
\begin{eqnarray}
&C_1P_{n_0}+C_1\sum_{n\le n_0-1} e^{-\mu ((n_0-n)\delta r -2\delta r/3)} P_{n}\le C_1 \sum_{n\le n_0} e^{-\mu (n_0-n)\delta r/3} P_{n}. \label{1564Inequality_for_first}
\end{eqnarray}
By the use of these inequalities~\eqref{1564Inequality_for_first00} and \eqref{1564Inequality_for_first}, the inequality~\eqref{inequality_for_P_n_n_0_p1} reduces to
\begin{eqnarray}
P_{n\ge n_0+1} \Biggl[ 2s \delta E_0 \biggl(\frac{\delta r}{3}\biggr)^2-  C_1e^{-\mu \delta r/3}  \Biggr]   &\le  C_1 \sum_{n\le n_0} e^{-\mu (n_0-n)\delta r/3} P_{n},\label{1564Inequality_for_first2}
\end{eqnarray}
which is generalized  to  the inequality~\eqref{General_inequality_P_n_01} with
\begin{eqnarray}
A= \frac{ C_1}{ 2s \delta E_0 (\delta r/3)^2-  C_1e^{-\mu \delta r/3}} ,\quad B=e^{-\mu \delta r/3}. \label{choice_of_A_and_B}
\end{eqnarray}
We now have to choose the parameter $\delta r$ appropriately in order to derive the main inequality.

In the following, we prove that if we choose $\delta r$ as $\xi_1$ in Eq.~\eqref{the_way_of_choice_of_xi}, the coefficients $A$ and $B$ satisfy the inequalities 
\begin{eqnarray}
A\le \frac{1}{2e}\quad  {\rm and}  \quad B\le \frac{1}{2e},   \label{Condition_for_A_and_B}
\end{eqnarray}
respectively.
For $A\le 1/(2e)$, we derive the following from Eq.~\eqref{choice_of_A_and_B}:
\begin{eqnarray}
&2eC_1 \le 2s \delta E_0 \biggl(\frac{\delta r}{3}\biggr)^2 - C_1e^{-\mu \delta r/3} \le 2s \delta E_0 \biggl(\frac{\delta r}{3}\biggr)^2-\frac{C_1}{2e} \nonumber \\
&\delta r \ge \frac{3}{2} \sqrt{\frac{(4e^2+1)C_1}{e s\delta E_0}},
\end{eqnarray}
where we utilized the inequality $e^{-\mu \delta r/3} = B \le 1/(2e)$.
For $B \le 1/(2e)$, we obtain
\begin{eqnarray}
\delta r \ge  \frac{3\ln (2e)}{\mu}.
\end{eqnarray} 
Therefore, if we choose $\delta r$ as $\xi_1= \max \biggl(\frac{3}{2} \sqrt{\frac{(4e^2+1)C_1}{e s \delta E_0}}, \frac{3\ln (2e)}{\mu}\biggr) $, we obtain the inequality~\eqref{Condition_for_A_and_B}.

We now have the inequality~\eqref{Case_1_main_inequality34} in the case~1 and the inequality~\eqref{General_inequality_P_n_01} with \eqref{Condition_for_A_and_B} in the case~2 by letting $\delta r=\xi_1$; the inequality in the case~2 is given by
\begin{eqnarray}
P_{n\ge n_0+1} \le \frac{1}{2e} \biggl[P_{n_0} + \frac{P_{n_0-1}}{2e} + \frac{P_{n_0-2}}{(2e)^2}+ \cdots \frac{P_{1}}{(2e)^{n_0-1}}   \biggr]. \label{General_inequality_P_n_01_A_B_determined}
\end{eqnarray}
Because the inequality~\eqref{Case_1_main_inequality34} is stronger than the inequality~\eqref{General_inequality_P_n_01_A_B_determined},
we only have to consider the inequality~\eqref{General_inequality_P_n_01_A_B_determined}.

We then prove by induction that the inequality~\eqref{General_inequality_P_n_01} reduces to 
\begin{eqnarray}
P_{n\ge n_0+1} \le A(A+B)^{n_0-2}(P_2+BP_1). \label{The_inequality_want_to_prove}
\end{eqnarray}
Note that we consider the case $n_0\ge2$.
For $n_0=2$, the inequality~\eqref{General_inequality_P_n_01} directly gives  the inequality~\eqref{The_inequality_want_to_prove}. 
Next, we prove the case of $n_0=\tilde{n}_0$ under the assumption that the inequality is satisfied for $n_0\le \tilde{n}_0-1$.
First, we have 
\begin{eqnarray}
P_{n_0+1}  \le P_{n\ge n_0+1}  \le  A(A+B)^{n_0-2}(P_2+BP_1)
\end{eqnarray}
for $n_0=2,3,\ldots, \tilde{n}_0-1$.
We then calculate the case of $n_0=\tilde{n}_0$ as
\begin{eqnarray}
P_{n\ge \tilde{n}_0+1} &\le  A (P_{\tilde{n}_0} + B P_{\tilde{n}_0-1}+ B^2 P_{\tilde{n}_0-2}+ \cdots B^{\tilde{n}_0-1} P_{1} )\nonumber \\
&\le  AB^{\tilde{n}_0-2} (P_2+BP_1) +    A^2 (P_2+BP_1) \sum_{k=3}^{\tilde{n}_0} (A+B)^{k-3} B^{\tilde{n}_0-k}  \nonumber \\
&=A (P_2+BP_1) \Biggl\{ B^{\tilde{n}_0-2}  +   B^{\tilde{n}_0-2} \biggl[\Bigl(\frac{A+B}{B}\Bigr)^{\tilde{n}_0-2} -1  \biggr]\Biggr\} \nonumber \\
&=A(A+B)^{\tilde{n}_0-2}(P_2+BP_1).
\end{eqnarray}
This completes the proof of the inequality~\eqref{The_inequality_want_to_prove}.

Using the inequality~\eqref{The_inequality_want_to_prove}, we reduce the inequality~\eqref{General_inequality_P_n_01_A_B_determined} to
\begin{eqnarray}
\fl P_{n\ge n_0+1} \le  A (A +B )^{n_0-2}(P_2+BP_1) \le \frac{P_2+P_1/(2e)}{2}e^{-n_0+1}\le \frac{2e(2-s)+1}{4(2e+1)}e^{-n_0}, \label{Inequality_P_n0_plus_1_exp}
\end{eqnarray}
where we utilized the inequality~\eqref{1Definition_of_P_0}, which gives $P_2\le (1-s)/(2e+1)$, and have $P_2+P_1/(2e) \le  (1-s)/(2e+1) + 1/(2e)= \bigl[2e(2-s)+1)\bigr]/\bigl[2e(2e+1)\bigr]$.
From the inequality~\eqref{Inequality_P_n0_plus_1_exp}, we obtain
\begin{eqnarray}
P_{n\ge N} \le \frac{2e(2-s)+1}{4(2e+1)} e^{- N+1} ,
\end{eqnarray}
where $N$ is defined in Eq.~\eqref{The_Definition_of_Large_N}.
Because we have
\begin{eqnarray}
N\equiv \biggl \lceil \frac{R-r_1}{\xi_1}     \biggr \rceil +1    \ge    \frac{R-r_1}{\xi_1} +1
\end{eqnarray}
in the case of $\delta r=\xi_1$, we finally obtain
\begin{eqnarray}
P_{n\ge N} \le  \frac{2e(2-s)+1}{4(2e+1)} \exp \Biggl(-\frac{R-r_1}{\xi_1}  \Biggr) . \label{FInal_inequality_ready}
\end{eqnarray}
Thus, by applying the inequality~\eqref{P_x_R_and_P_n_N} to \eqref{FInal_inequality_ready}, we prove the main inequality~\eqref{Final_form_of_the_inequality_decay}.  $\opensquare$


We have so far considered the case where the hopping rate decays exponentially as in the inequality~\eqref{Condition_for_interaction}.
If we  consider the nearest-neighbor hopping, namely,  
\begin{eqnarray}
&\biggl \|\sum_{i,j=1}^{N_0}  (  h_{(x,i),(x',j)} a_{(x,i)}^{\dagger} a_{(x',j)} + {\rm h.c.}    ) \biggr \|_2  
\cases{ \le \mathcal{V}_0  \quad   {\rm for} \quad |x-x'|= 1   ,     \\
  = 0  \quad   {\rm for} \quad  |x-x'|\ge 2   ,}   \label{Condition_for_interaction2}
\end{eqnarray}
we can obtain a stronger upper bound than that of Theorem~1.

\textit{Theorem~2}.
When we assume 
\begin{eqnarray}
R \ge r_1 \equiv  \sqrt{\frac{e+1}{1-s}} \Delta X \quad {\rm with} \quad 0<s<1,  \label{lower_bound_R2}
\end{eqnarray}
the distribution $P(|x-\langle x \rangle| \ge  R )$ satisfies the following inequality:
\begin{eqnarray}
P(|x-\langle x \rangle| \ge  R ) \le \frac{e(1-s)}{e+1}  \exp \Biggl(-\frac{R-r_1}{\xi_2} \Biggr) , \label{Final_form_of_the_inequality_decay2}
\end{eqnarray}
where
\begin{eqnarray}
\xi_2 \equiv  \sqrt{\frac{ e\mathcal{V}_0 }{s \delta E_0}}+2       .\label{the_way_of_choice_of_xi2}
\end{eqnarray}

\textit{Proof}.

The proof is similar to that of Theorem~1 and we can prove Theorem~2 by following the proof of Theorem~1 though the details are different.  
As in the proof of Theorem~1 (Figure~1), we split the coordinate space into some regions:
\begin{eqnarray} 
r_0=0,\quad r_1=\sqrt{\frac{e+1}{1-s}} \Delta X, \quad  {\rm and} \quad   r_n=r_{n-1} +\delta r \quad (n\ge 2) .      \label{each_region_definition_rn2}
\end{eqnarray} 

We also define the region $N$ as the one which includes the point $x=R$; the definition is given in Eq.~\eqref{The_Definition_of_Large_N}.
The point $x=R$ then satisfies $r_{N-1} \le R <  r_N$ and $N$ is larger than $2$ because of $R \ge r_1$.
We here define $P_{n}$ and $P_{n\ge n_0}$ as in Eqs.~\eqref{THe_Definition_of_Large_P_n} and \eqref{THe_Definition_of_Large_P_n_ge_n_0}.
We then consider the upper bound of $P_{n\ge N} $ instead of $P(|x-\langle x \rangle| \ge  R )$ because of the inequality~\eqref{P_x_R_and_P_n_N}.

 \begin{figure}
\centering
\includegraphics[clip, scale=0.6]{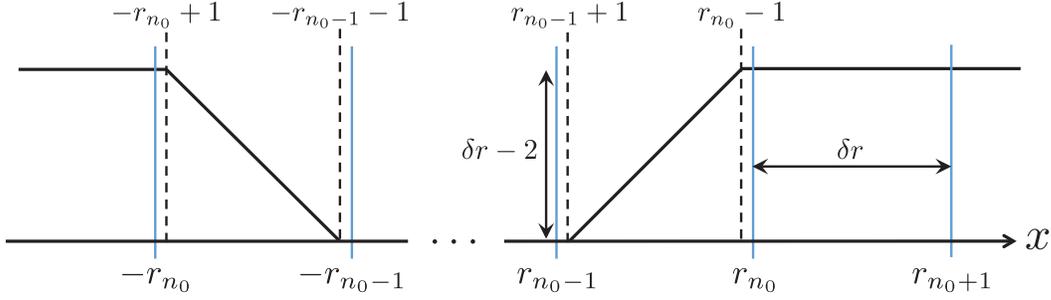}
\caption{The function $g(x)$ in operator $G$, which is given by Eq.~\eqref{the_choice_of_g_x_in_theoREm_2}.
}
\label{fig:definition_of_the_function_gx2}
\end{figure}

We here choose the function $g(x)$ as follows (Figure.~\ref{fig:definition_of_the_function_gx2}):
\begin{eqnarray}
\fl&g(x) = 
\cases{ 0\quad   {\rm for} \quad  |x|\le r_{n_0-1}+1,      \\
|x| -(r_{n_0-1}+1 ) \quad  {\rm for} \quad   r_{n_0-1}+1  \le |x| \le   r_{n_0}-1  , \\  
  \delta r -2  \quad   {\rm for} \quad |x| \ge   r_{n_0}-1  .}   \label{the_choice_of_g_x_in_theoREm_2}
\end{eqnarray}
From the above function $g(x)$, we calculate the lower bound of $\bra{\psi_0}G^2\ket{\psi_0}$ and the upper bounds of $|\bra{\psi_0} H_{{\rm OD}} \ket{\psi_0}|$ and $\bra{\psi_0}G\ket{\psi_0}^2$.

We first obtain the upper bound of $|\bra{\psi_0} H_{{\rm OD}} \ket{\psi_0}|$.
Now, $H_{{\rm OD}}$ is given by
\begin{eqnarray}
H_{{\rm OD}} &= \sum_{x} |g(x)-g(x-1)|^2 \sum_{i, j=1}^{N_0} (h_{(x,i),(x-1,j)}  a_{x,i}^\dagger a_{x-1,j} +{\rm h.c.} ),
\end{eqnarray}
and we have 
\begin{eqnarray}
& g(x)-g(x-1) 
\cases{ \le 1 \quad   {\rm for} \quad  r_{n_0-1}+1  \le |x| \le   r_{n_0}  ,    \\
 =0 \quad  {\rm  otherwise}.}  
\end{eqnarray}
Therefore, we obtain
\begin{eqnarray}
\fl | \bra{\psi_0} H_{{\rm OD}}   \ket{\psi_0}| &= \sum_{x} |g(x)-g(x-1)|^2 \sum_{i, j=1}^{N_0}  (h_{(x,i),(x-1,j)}  \alpha_{x,i}^\dagger \alpha_{x-1,j} +{\rm c.c.} )  \nonumber \\
\fl&\le \sum_{r_{n_0-1}+1  \le |x| \le   r_{n_0}}    \biggl \|\sum_{i,j=1}^{N_0} (  h_{(x,i),(x-1,j)} a_{(x,i)}^{\dagger} a_{(x-1,j)} + {\rm h.c.}    ) \biggr \|_2  (p_{x}+ p_{x-1})  \nonumber \\ 
\fl &\le  \sum_{r_{n_0-1}+1  \le |x| \le   r_{n_0}}  \mathcal{V}_0  (p_{x}+ p_{x-1}) \le 2 \mathcal{V}_0  P_{n_0}      . \label{last_form_of_the_inequality_lemma222}
\end{eqnarray}

Next, we calculate the upper bound of  $\bra{\psi_0}G\ket{\psi_0}^2$ and the lower bound of $\bra{\psi_0}G^2\ket{\psi_0}$ as 
\begin{eqnarray}
\bra{\psi_0}G\ket{\psi_0}^2 \le  (P_{n_0}+P_{n\ge n_0+1})^2 ( \delta r-2 )^2 \label{inequality_F_square_22}
\end{eqnarray} 
and 
\begin{eqnarray}
\bra{\psi_0}G^2\ket{\psi_0}\ge P_{n\ge n_0+1}  ( \delta r-2 )^2 ,\label{inequality_F_square2}
\end{eqnarray} 
respectively.

In the following, we consider the two cases 1 and 2.\\
\textit{(case~1)}
\begin{eqnarray}
(1-s)  \bra{\psi_0}G^2\ket{\psi_0}\le \bra{\psi_0}G\ket{\psi_0}^2 .  \label{G2_AG2_condition_case12}
\end{eqnarray}
\textit{(case~2)}
\begin{eqnarray}
(1-s)  \bra{\psi_0}G^2\ket{\psi_0}>  \bra{\psi_0}G\ket{\psi_0}^2  . \label{G2_AG2_Condition_case22}
\end{eqnarray}

We first consider the case~1.
From the inequalities~\eqref{inequality_F_square_22} and \eqref{inequality_F_square2}, we  obtain
\begin{eqnarray}
\fl P_{n\ge n_0+1} ( \delta r-2 )^2\le \bra{\psi_0}G^2\ket{\psi_0}\le \frac{1}{1-s} \bra{\psi_0}G\ket{\psi_0}^2   \le \frac{ ( \delta r-2 )^2}{1-s}(P_{n_0}+P_{n\ge n_0+1})^2,
\end{eqnarray}
which reduces to
\begin{eqnarray}
P_{n\ge n_0+1}\le  \frac{1}{1-s}(P_{n_0}+P_{n\ge n_0+1})^2.
\end{eqnarray}
We then utilize the Chebyshev inequality as 
\begin{eqnarray}
P_{n_0}+P_{n\ge n_0+1}  \le  \frac{(\Delta X)^2}{r_1^2} =\frac{1-s}{e+1}  \equiv (1-s) P_{0}, \label{1Definition_of_P_02}
\end{eqnarray}
because we have $|x| \ge r_1=\sqrt{\frac{e+1}{1-s}}\Delta X$ in the $n_0$th region ($n_0\ge 2$).
Therefore, we have
\begin{eqnarray}
P_{n\ge n_0+1}   & \le P_0 (P_{n_0}+P_{n\ge n_0+1}) \nonumber \\
 P_{n\ge n_0+1} & \le \frac{P_0}{1-P_0}P_{n_0} = \frac{1}{e} P_{n_0},   \label{Case_1_main_inequality342}
\end{eqnarray}
where $P_{0}=1/(e+1)$ from the definition~\eqref{1Definition_of_P_02}.

We next consider the case~2.
From the inequalities~\eqref{inequality_F_square2} and \eqref{G2_AG2_Condition_case22}, we obtain
\begin{eqnarray}
\fl (\Delta G)^2= \bra{\psi_0}G^2\ket{\psi_0} - \bra{\psi_0}G\ket{\psi_0}^2  > s \bra{\psi_0}G^2\ket{\psi_0} \ge s P_{n\ge n_0+1} ( \delta r-2 )^2 .\label{last_form_of_the_inequality2_F2}
\end{eqnarray}
From the inequalities \eqref{last_form_of_the_inequality_lemma222} and \eqref{last_form_of_the_inequality2_F2}, we reduce the inequality~\eqref{New_Final_form_of_the_inequality} to
\begin{eqnarray}
2s P_{n\ge n_0+1} \delta E_0  ( \delta r-2 )^2  \le   2\mathcal{V}_0  P_{n_0} , 
\end{eqnarray}
which further reduces to 
\begin{eqnarray}
P_{n\ge n_0+1} \le  \frac{ \mathcal{V}_0}{s  \delta E_0  ( \delta r-2 )^2}  P_{n_0} .  \label{inequality_for_P_n_n_0_p12}
\end{eqnarray}
If we choose $\delta r= \xi_2$, we obtain the inequality~\eqref{Case_1_main_inequality342} in the case~2 as well.

We therefore obtain the inequality~\eqref{Case_1_main_inequality342} in both cases.
By applying the inequality~\eqref{Case_1_main_inequality342} iteratively, we obtain
\begin{eqnarray}
P_{n\ge N} \le  eP_2 e^{- N+1} \le \frac{e(1-s)}{e+1} e^{- N+1},
\end{eqnarray}
where we utilized the inequality~\eqref{1Definition_of_P_02}, which gives $P_2\le (1-s)/(e+1)$.
Because we have
\begin{eqnarray}
N\equiv \biggl \lceil \frac{R-r_1}{\xi_2}     \biggr \rceil +1    \ge    \frac{R-r_1}{\xi_2} +1
\end{eqnarray}
in the case of $\delta r=\xi_2$, we finally obtain
\begin{eqnarray}
P(|x-\langle x \rangle| \ge  R )  \le P_{n\ge N} \le \frac{e(1-s)}{e+1}  \exp \Biggl(-\frac{R-r_1}{\xi_2}  \Biggr) . \label{FInal_inequality_ready2}
\end{eqnarray}
Thus, we prove the main inequality~\eqref{Final_form_of_the_inequality_decay2}.  $\opensquare$


\subsection{Tightness of the main inequalities}
We here discuss the tightness of the main inequalities~\eqref{Final_form_of_the_inequality_decay} and \eqref{Final_form_of_the_inequality_decay2}.
For this purpose, we compare the upper bounds of the localization length in the following Hamiltonian:
\begin{eqnarray}
 H=\sum_{x=-L/2}^{L/2}  ( h_{x,x'} a_{x}^{\dagger} a_{x'} + {\rm h.c.}    ) +h_0 a_{0}^{\dagger} a_{0} \label{Hamiltonian_tight_binding_One_Dimensional}
\end{eqnarray}
with 
\begin{eqnarray}
h_{x,x'}\le e^{-|x-x'|}  
\quad {\rm and} \quad 
h_{x,x'}  
\cases{ =1  \quad   {\rm for} \quad |x-x'|=1   ,     \\
  = 0  \quad   {\rm for} \quad  |x-x'|\ge 2   } ,  \label{numerical_comparation_LOlength_case_2}
\end{eqnarray}
where $C_v=1$ and $\mu=1$ in the inequality~\eqref{Condition_for_interaction} and $\mathcal{V}_0=1$ in the inequality~\eqref{Condition_for_interaction2}, respectively.
In this model, there is a defect at the point $x=0$ and the distribution $p_x$ in Eq.~\eqref{definition_of_the_density} approximately decays as
\begin{eqnarray}
p_x \propto e^{-|x|/\xi}   \quad  {\rm with } \quad \xi= \frac{\Delta X}{\sqrt{2} }. \label{Ap_p_x_beHavior}
\end{eqnarray}
For the Hamiltonian with the nearest-neighbor hopping, this approximation can be shown to be exact in the limit of $L\to \infty$~\cite{Apollaro}.
 
We, in the following, compare $\Delta X/\sqrt{2} $ with $\xi_1$ and $\xi_2$ in Eqs.~\eqref{the_way_of_choice_of_xi} and \eqref{the_way_of_choice_of_xi2}; we calculate 
\begin{eqnarray}
\frac{\sqrt{2} \xi_1}{\Delta X} \quad {\rm and} \quad  \frac{\sqrt{2}  \xi_2}{\Delta X} \label{two_plot_dAta_}
\end{eqnarray}
for $-1 \le h_0 \le -0.01$ with $L=500$, where we take the parameter $s$ as $1/2$. 
The reason why we take $h_0 \le -0.01$ is that the approximation of \eqref{Ap_p_x_beHavior} is not good in the limit of $h_0\to 0$, where $p_x\sim \cos(\pi |x| /L)$.

\begin{figure}
\centering
\subfigure[]{
\includegraphics[clip, scale=0.7]{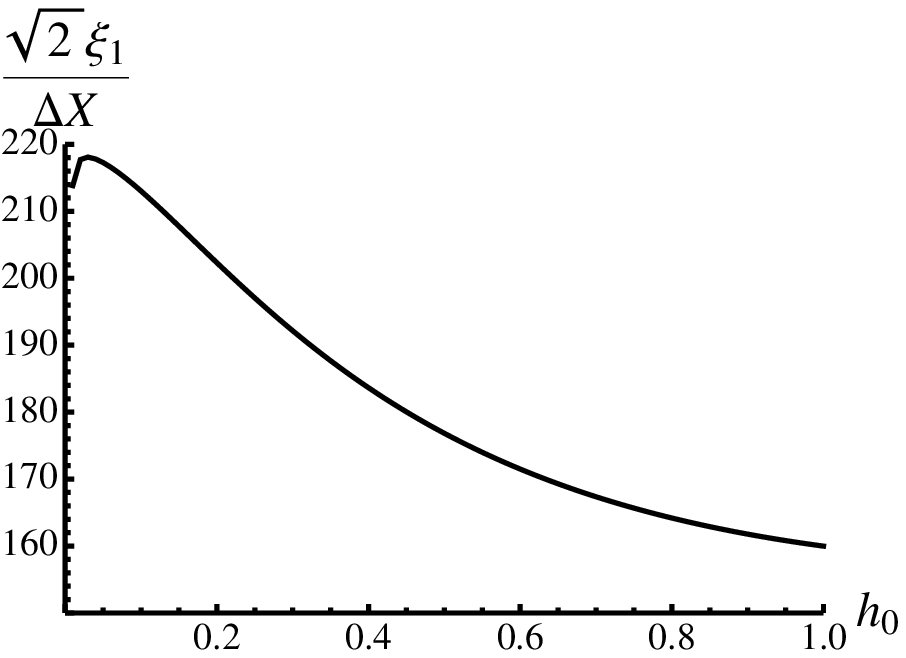}
}
\subfigure[]{
\includegraphics[clip, scale=0.7]{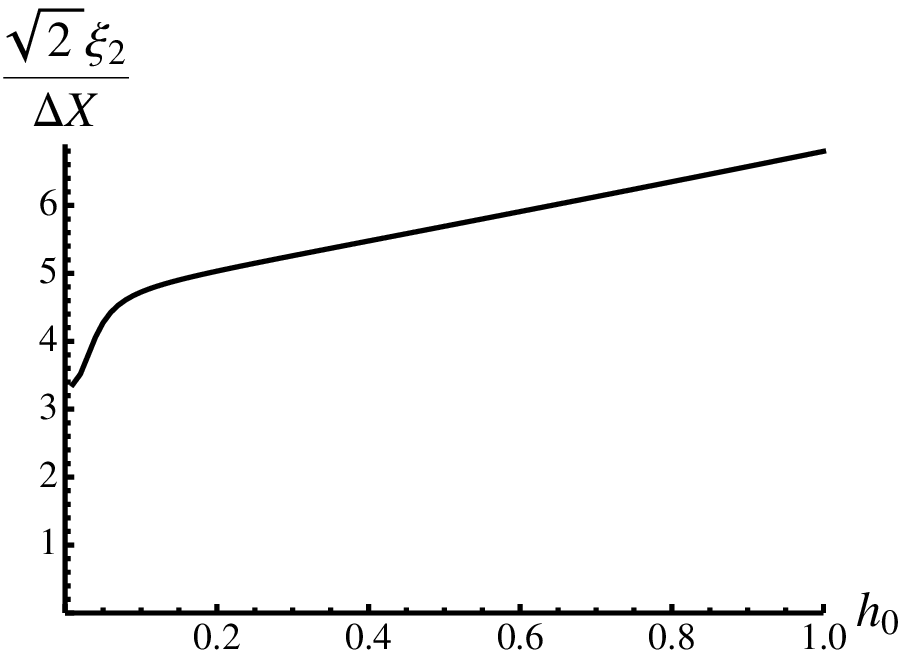}
}
\caption{The plots of (a) $\sqrt{2}\xi_1/\Delta X$  and  (b) $\sqrt{2}\xi_2/\Delta X$ with respect to $h_0$ in Eq.~\eqref{Hamiltonian_tight_binding_One_Dimensional}.
We consider $-1 \le h_0 \le -0.01$ and $L=500$ in Eq.~\eqref{Hamiltonian_tight_binding_One_Dimensional} and $s=1/2$ in Eqs.~\eqref{the_way_of_choice_of_xi} and \eqref{the_way_of_choice_of_xi2}.

}
\label{fig:Compare_xi_delta_x}
\end{figure}

In Figure~\ref{fig:Compare_xi_delta_x}, we show the numerical plots of  \eqref{two_plot_dAta_} in order to discuss the tightness of the localization lengths $\xi_1$ and $\xi_2$.
We ensure that the tightness of $\xi_2$ is much better than that of $\xi_1$. 
As for the tightness of $\xi_1$, we also have a possibility to refine the present upper bound to some extent.
For example, we have a degree of freedom how to choose the function $g(x)$ of the operator $G$.

\section{Conclusion}
We have given the rigorous proof to the empirical inference~$\xi \sim \delta E_0^{-1/2}$ and obtained  the inequality~$\xi \le \const. \times \delta E_0^{-1/2}$ in general one-particle systems with short-range hopping.
For the proof, we utilized the inequality~\eqref{New_Final_form_of_the_inequality} which characterize the complementary relationship between the spectral gap and the fluctuation.
In this inequality, we have a degree of freedom how to choose the operator $G$, which is defined in Eq.~\eqref{G_tight_binding}.
We can extract the local properties of the distribution function $p_x$ by choosing the operator $G$ as in Eq.~\eqref{choice_of_g_x_of_the_OpeRator_G}.
By putting together such local properties, we obtain the main inequality~\eqref{Final_form_of_the_inequality_decay}.
We have obtained a stronger upper bound~\eqref{Final_form_of_the_inequality_decay2} for a Hamiltonian with nearest-neighbor hopping.
The point of the proof is the choice of the operator $G$.

We have also tested the tightness of these upper bounds in the tight-binding Hamiltonian with a diagonal defect.
In this model, we can easily define the localization length and compare it to the upper bound.
We tested two upper bounds given in Eqs.~\eqref{the_way_of_choice_of_xi} and \eqref{the_way_of_choice_of_xi2}; they are given in the Hamiltonian with general short-range hopping~\eqref{Condition_for_interaction} and the Hamiltonian with nearest-neighbor hopping~\eqref{Condition_for_interaction2}, respectively.
We have ensured that the upper bound is tight in the case of the nearest-neighbor hopping.
The further refinement is a future problem, but we consider that it will be possible by choosing the operator $G$ more appropriately. 

In conclusion, we have refined the exponential clustering in the one-particle systems.
The point is that we have utilized the complementary inequality~\eqref{New_Final_form_of_the_inequality} instead of the Lieb-Robinson bound.
This fact indicates that the causality of the systems is not enough to characterize local properties of the ground state. 
Our complementary relationship would influence the fundamental properties of the ground state in different ways from the Lieb-Robinson bound, although our proof is now applicable only to the one-particle systems.
We plan to extend the present theory to more general systems as a future problem.

\section*{ACKNOWLEDGMENT}
The present author is grateful to Professor Naomichi Hatano for helpful discussions and comments.
This work was supported by the Program for Leading Graduate 
Schools, MEXT, Japan.

\appendix

\section{Derivation of Eq.~\eqref{definition_of_HOD}} \label{appendixA}
We here derive Eq.~\eqref{definition_of_HOD}. 
From the result in Ref.~\cite{Kuwahara}, we can obtain
\begin{eqnarray}
\delta E_0 \le \frac{\bra{\psi_0} G H G \ket{\psi_0} -\bra{\psi_0} G^2 \ket{\psi_0}E_0}{(\Delta G)^2}.
\end{eqnarray}
We can simplify this inequality as
\begin{eqnarray}
\delta E_0 \le \frac{\bra{\psi_0}  H_{\rm OD}  \ket{\psi_0}}{2(\Delta G)^2}  \equiv \frac{ -\bra{\psi_0} [G, [G,H]] \ket{\psi_0}}{2(\Delta G)^2}, \label{express_H_OD_1}
\end{eqnarray}
because
\begin{eqnarray}
& \bra{\psi_0} G H G \ket{\psi_0} -\bra{\psi_0} G^2 \ket{\psi_0}E_0 \nonumber \\
=& \bra{\psi_0} G [H, G] \ket{\psi_0} = \bra{\psi_0}[ G ,H] G \ket{\psi_0}= \frac{ -\bra{\psi_0} [G, [G,H]] \ket{\psi_0}}{2},
\end{eqnarray}
where we utilized $H \ket{\psi_0}=E_0 \ket{\psi_0}$.
Therefore, we obtain the form of $H_{\rm OD}$ as 
\begin{eqnarray}
 H_{\rm OD} = -\bra{\psi_0} [G, [G,H]] \ket{\psi_0}. \label{express_H_OD_2}
\end{eqnarray}
By applying the definition of $H$ and $G$ to Eq.~\eqref{express_H_OD_2}, we obtain Eq.~\eqref{definition_of_HOD} after straightforward algebra.

\section{Proof of the inequality~\eqref{last_form_of_the_inequality_lemma}} \label{appendixB}

We first calculate $| \bra{\psi_0} H_{{\rm OD}}   \ket{\psi_0}|$ as in \eqref{Inequality_chebyshef}:
\begin{eqnarray}
| \bra{\psi_0} H_{{\rm OD}}   \ket{\psi_0}| &= \sum_{x, x'} |g(x)-g(x')|^2 \sum_{i, j=1}^{N_0} (h_{(x,i),(x',j)}  \alpha_{(x,i)}^\ast \alpha_{x',j} +{\rm c.c.} )  \nonumber \\
&\le  2\sum_{x, x'}|g(x)-g(x')|^2\mathcal{V}(x-x') p_{x} \equiv  \sum_{x}\mathcal{V}_{g,x} p_x      . \label{last_form_of_the_equality_lemma}
\end{eqnarray}
From the inequality \eqref{Condition_for_interaction}, we can calculate the upper bound of $\mathcal{V}_{g,x}$.
In the following, we calculate it in the case $x\ge 0$, but the same calculation can be applied to the case $x\le 0$.

In the case $0 \le x \le  r_{n_0-1}+ \delta r/3$, we have $g(x)=0$ and $|g(x)-g(x')|^2=|g(x')|^2$, which is followed by
\begin{eqnarray}
\fl \mathcal{V}_{g,x}&=2 \sum_{x'} |g(x')|^2\mathcal{V}(x-x')  \nonumber \\
\fl &=  2\sum_{r_{n_0-1}+ \delta r/3  \le |x'|\le r_{n_0-1}+ 2\delta r/3} \bigl[|x'| -(r_{n_0-1}+ \delta r/3 )\bigr]^2\mathcal{V}(x-x') \nonumber \\
\fl &\quad + 2\sum_{|x'|\ge r_{n_0-1}+ 2\delta r/3} (\delta r/3 )^2\mathcal{V}(x-x')   \nonumber \\
\fl &\le   4\sum_{r_{n_0-1}+ \delta r/3  \le x'\le r_{n_0-1}+ 2\delta r/3} \bigl[|x'| -(r_{n_0-1}+ \delta r/3 )\bigr]^2\mathcal{V}(x-x') \nonumber \\
\fl &\quad + 4\sum_{x'\ge r_{n_0-1}+ 2\delta r/3} (\delta r/3 )^2\mathcal{V}(x-x')   \nonumber \\
\fl &= 4C_v e^{-\mu (r_{n_0-1}+ \delta r/3-|x|)} \Biggl [ \sum_{0\le x' \le \delta r/3} (x')^2 e^{-\mu x'}+ \sum_{x' \ge \delta r/3} (\delta r/3)^2 e^{-\mu x'}\Biggr] \nonumber \\
\fl &\le 4C_v e^{-\mu (r_{n_0-1}+ \delta r/3-|x|)} \sum_{0\le x'\le \infty} (x')^2 e^{-\mu x'}\le C_1 e^{-\mu (r_{n_0-1}+ \delta r/3-|x|)}, \label{last_form_of_the_inequality_lemma1}
\end{eqnarray}
where we utilized the definition of $C_1$ in Eq.~\eqref{the_definition_of_C1} and  the inequalities
\begin{eqnarray}
\mathcal{V}(x-x') \le \mathcal{V}(x-|x'|)\quad {\rm for} \quad x\ge 0 \ \ {\rm and}\ \  x'\le 0,
\end{eqnarray}
from the second line to the third line and
\begin{eqnarray}
\sum_{0\le x' \le \infty} (x')^2 e^{-\mu x'} \le  \sum_{x'=0}^{\infty} (x'+1)^2 e^{-\mu x'},
\end{eqnarray}
from the fourth line to the fifth line, respectively.

In the similar way, we calculate the case $x\ge  r_{n_0-1}+ 2\delta r/3$, where $g(x)=\delta r/3$:
\begin{eqnarray}
\fl\mathcal{V}_{g,x}&= 2\sum_{x'}|\delta r/3-g(x')|^2\mathcal{V}(x-x')  \nonumber \\
\fl&=2 \sum_{r_{n_0-1}+ \delta r/3  \le |x'|\le r_{n_0-1}+ 2\delta r/3} \bigl[-|x'|+(r_{n_0-1}+2 \delta r/3 )\bigr]^2\mathcal{V}(x-x') \nonumber \\
\fl&\quad + 2\sum_{|x'|\le r_{n_0-1}+ \delta r/3} ( \delta r/3 )^2\mathcal{V}(x-x')   \nonumber \\
\fl&\le 4C_v e^{-\mu (|x|-r_{n_0-1}- 2\delta r/3)} \Biggl [  \sum_{0\le x' \le \delta r/3} (x')^2 e^{-\mu x'}+ \sum_{x' \ge \delta r/3} (\delta r/3)^2 e^{-\mu x'} \Biggr ]  \nonumber\\
\fl&\le 4C_v e^{-\mu (|x|-r_{n_0-1}- 2\delta r/3)} \sum_{0\le x'\le \infty} (x')^2 e^{-\mu  x'}\le C_1 e^{-\mu (|x|-r_{n_0-1}- 2\delta r/3)}.\label{last_form_of_the_inequality_lemma2}
\end{eqnarray}
Finally, we consider the case $r_{n_0-1}+\delta r/3\le x \le  r_{n_0-1}+2 \delta r/3$ as 
\begin{eqnarray}
\mathcal{V}_{g,x}&=2 \sum_{x'} |g(x)-g(x')|^2\mathcal{V}(x-x')\nonumber \\
&=2 \sum_{x'} \bigl[ x -(r_{n_0-1}+ \delta r/3 )-g(x')\bigr] ^2\mathcal{V}(x-x') 
\end{eqnarray}
and 
\begin{eqnarray}
\bigl[ x -(r_{n_0-1}+ \delta r/3 )-g(x')\bigr] ^2\le (x-x')^2
\end{eqnarray}
for arbitrary $x'$.
Therefore, we calculate $\mathcal{V}_{g,x}$ as
\begin{eqnarray}
\mathcal{V}_{g,x}&\le 2C_v \sum_{x' \le x} (x-x')^2 e^{-\mu |x-x'|}+2 C_v \sum_{x' \ge x} (x-x')^2 e^{-\mu |x-x'|}\le   C_1.\label{last_form_of_the_inequality_lemma3}
\end{eqnarray}
We summarize the inequalities \eqref{last_form_of_the_inequality_lemma1}, \eqref{last_form_of_the_inequality_lemma2} and \eqref{last_form_of_the_inequality_lemma3}:
\begin{eqnarray}
\fl&\mathcal{V}_{g,x} \le  
\cases{ C_1 e^{-\mu (r_{n_0-1}+ \delta r/3-|x|)} \quad   {\rm for} \quad  |x|  \le r_{n_0-1}+ \delta r/3,     \\
C_1 \quad  {\rm for} \quad  r_{n_0-1}+ \delta r/3 \le |x|\le r_{n_0-1} + 2\delta r/3 , \\  
  C_1 e^{-\mu (|x|-r_{n_0-1}- 2\delta r/3)} \quad   {\rm for} \quad   |x| \ge r_{n_0-1} + 2\delta r/3  .}  
\end{eqnarray}
From the inequality \eqref{last_form_of_the_equality_lemma}, we obtain the inequality \eqref{last_form_of_the_inequality_lemma}.
This completes the proof.


\section*{References}

\renewcommand{\refname}{\vspace{-1cm}}



\begin{thebibliography}{00}

\bibitem{LR_bound} 
E. H. Lieb and D. W. Robinson, Commun. Math. Phys. {\bf28}, 251 (1972).





\bibitem{Nachtergaele3}     
B. Nachtergaele, Y. Ogata and R. Sims, J. Stat. Phys. {\bf124}, 1 (2006).



\bibitem{Bravyi2}     
S. Bravyi, M. B. Hastings and F. Verstraete Phys. Rev. Lett, {\bf97}, 050401 (2006). 
 

\bibitem{Osborne}  
C. K. Burrell and T. J. Osborne, Phys. Rev. Lett. {\bf99}, 167201 (2007).

\bibitem{Hamma} 
A. Hamma, F. Markopoulou,  I. Pr{\'e}mont-Schwarz and S. Severini, Phys. Rev. Lett. {\bf102},  017204 (2009).



\bibitem{Nachtergaele4} 
B. Nachtergaele, H. Raz, B. Schlein and R. Sims, Commun. Math. Phys. {\bf286}, 1073 (2009).




\bibitem{Schwarz}    
 I. Pr{\'e}mont-Schwarz and  J. Hnybida, Phys. Rev. A {\bf81}, 062107 (2010).

\bibitem{Cheneau}  
M. Cheneau, P. Barmettler, D. Poletti, M. Endres, P. Schauss, T. Fukuhara, C. Gross, I. Bloch, C. Kollath and S. Kuhr, Nature {\bf481}, 484 (2012).







\bibitem{Hastings}   
M. B. Hastings, Phys. Rev. B. {\bf 69}, 104431 (2004).



\bibitem{Nachtergaele2}  
B. Nachtergaele and R. Sims, Commun. Math. Phys. {\bf276}, 437 (2007).


\bibitem{Hastings4} 
M. B. Hastings, J. Stat. Mech. P08024 (2007).

\bibitem{Hastings5}  
M. B. Hastings, Phys. Rev. B {\bf76}, 035114 (2007).

\bibitem{Eisert} 
 J. Eisert, M. Cramer and M. B. Plenio, Rev. Mod. Phys. {\bf82}, 277 (2010). 

\bibitem{Gottesman}      
D. Gottesman and M. B. Hastings, New J. Phys. {\bf12} 025002 (2010).



\bibitem{Hastings6}  
M. B. Hastings, arXiv: 1008.2337 [math-ph].



\bibitem{Hastings2}   
M. B. Hastings, Phys. Rev. Lett. {\bf93}, 140402 (2004).



\bibitem{Hastings3}   
M. B. Hastings and T. Koma, Commun. Math. Phys.  {\bf265}, 781 (2006).


\bibitem{Nachtergaele}  
B. Nachtergaele and R. Sims, Commun. Math. Phys. {\bf265}, 119 (2006).


\bibitem{Koma}  
T. Koma, J. Math. Phys. {\bf 48}, 023303 (2007).





\bibitem{Vojtan}    
M. Vojtan, Rep. Prog. Phys. {\bf66}, 2069 (2003).
  
\bibitem{BookQP}    
S. Sachdev, \textit{Quantum phase transitions} (Wiley Online Library, 2007).
  




\bibitem{Apollaro}    
T. J. G. Apollaro and F. Plastina, Phys. Rev. A {\bf74}, 062316 (2006).








\bibitem{Kuwahara} 
T. Kuwahara, J. Phys. A: Math. Theor. {\bf46} (2013).
















\end{thebibliography}
\end{document}